\documentclass[preprint,showpacs,preprintnumbers,amsmath,amssymb]{revtex4}
\usepackage{wrapfig}
\usepackage{graphicx}% Include figure files
\usepackage{dcolumn}% Align table columns on decimal point
\usepackage{bm}% bold math

\begin{document}

%\preprint{APS/123-QED}

\title{Self-tuning of threshold  for a two-state system}
\author{Boyoung Seo$^1$}

\author{Raishma Krishnan$^2$}

\author{Toyonori Munakata$^1$}
\email{munakata@amp.i.kyoto-u.ac.jp}

\affiliation{$^1$Department of Applied Mathematics and Physics, 
Graduate School of Informatics, Kyoto University, Kyoto 606-8501,
Japan\\
 $^2$Institute of Physics, Sachivalaya Marg, Bhubaneswar 751005,
Orissa, India}
%\date{\today}
\begin{abstract}
 A two-state  system (TSS) under time-periodic perturbations (to be 
regarded as input signals) 
is studied 
 in  connection with self-tuning (ST) of threshold and stochastic
 resonance (SR).
By ST, we observe the improvement of signal-to-noise
 ratio (SNR) in a weak noise region. Analytic approach to 
  a tuning equation  reveals that SNR improvement is  possible also for 
a large noise region and this is demonstrated  by Monte Carlo simulations of
 hopping processes in a TSS. ST and SR are discussed from 
a little more physical point of 
 energy transfer (dissipation) 
rate, which behaves in a similar way as SNR.
Finally ST is considered briefly for a double-well 
potential system (DWPS), which is closely related to the TSS.

\end{abstract}
\pacs{ 05.10.Gg, 02.70.Uu, 05.40.-a}% PACS, the Physics and Astronomy                             % Classification Scheme.
\maketitle

\section{INTRODUCTION}\label{sec:1}

Recently constructive or beneficial roles of noise gather
considerable interest in many fields, such as physical~\cite{1},
 and biological~\cite{2} sciences as well as engineering~\cite{3}.
One of the conspicuous effects of noise or random disturbance 
 is that it can drive a dynamical system out of an  equilibrium state,
thus giving a life time or Kramers time~\cite{4} to 
(metastable) equilibrium states. 

Simulated annealing method\cite{3}, which is used to search for  solutions to
minimization ( or more generally optimization) problems in a complex
system, employs noise to prevent 
a search   process from being trapped in  local minimum(metastable) 
 states.  Sophisticated algorithms are developed to efficiently 
escape from  local metastable states, which are  useful  for both  
simulated annealing and  efficient Monte Carlo simulations.\cite{5}   
  
Stochastic resonance(SR)\cite{1}, which stands for a phenomenon  in which
information transfer from  input  to output signals can be
significantly increased by  noise with appropriate (non-zero) intensity.
One can   comprehend SR  by  considering  a simple threshold system,\cite{6} 
which gives 1(0) as an output signal $x$ 
if input signal $s$ plus noise $\xi$ is larger(smaller) than a certain 
threshold value
$a$.  If an input signal $s$ is always smaller than $a$ and there is no
noise, $x$ is always equal to 0 and 
information transfer through the threshold system is impossible.
    By adding noise 
$\xi$ to $s$, there is  some possibility of $s+\xi> a$, producing
$x=1$ and 
information about $s$ is conveyed through the threshold system.
However   large noise deforms  original input signals too much, leading to 
no correlation between $s$ and $x$, resulting in no information transfer 
from  input to  output signals.  
 
As a  system similar to the threshold system mentioned above, %which is a little more complex than the
%simple threshold system mentioned above, 
let us consider    
an overdamped Brownian particle in a double-well potential driven
by  a sinusoidal time-periodic force, which  was  proposed and studied 
 as a model for 
Earth's ice ages\cite{7}.  This model   has an activation energy
%the difference in energy at  the local maximum and 
%the well bottom, 
 and the Gaussian Brownian noise $\xi_G$, 
which may be regarded as the threshold  value $a$ and 
the noise $\xi$, respectively in the threshold system. 
In this case information on  input signal, such as  the frequency 
$2\pi\omega_0$ of  
the sinusoidal  force,   is transfered as the peak position
in the power spectrum of  output signal. When the variance 
 ( or temperature from the fluctuation-dissipation theorem) of $\xi_G$ 
 is tuned to an appropriate value, which turns out to be non-zero, the
 signal-to-noise ratio(SNR) attains its  maximum value.

From this we may consider that SR has  a close relation with
 synchronization,
 especially when external disturbance is characterized by a  
frequency $f_0=2\pi\omega_0$.  In this regard we mention  
 stochastic synchronization, in which 
 an excitable system like neurons, responds in synchrony with external 
disturbance(signal), which also gathers lots of interest 
 in connection with   electroreceptors in the paddlefish\cite{8}.
                 
When  input signals are subthreshold,  
 ability of a threshold system to transfer information is 
considerably limited for weak noise, as mentioned above. 
 To improve information transfer in  this region, 
we proposed  recently a simple 
adaptation process\cite{9}  for a threshold value $a$
% in a simple threshold system, 
hinted by a self-tuning mechanism
proposed to explain auditory sensitivity\cite{10} when input signal 
becomes very weak. 

In this paper we consider effects of self-tuning(ST) of the threshold value 
for a two-state system(TSS) driven by a sinusoidal signal. 
One merit of TSS
   is that one can calculate  SNR  accurately\cite{11} 
by solving a differential equation, without 
 doing numerical experiments 
 to obtain the power spectrum, based on which SNR is 
usually calculated. In  Sec.~\ref{sec:2} we introduce our system, TSS and
a closely related double-well potential system(DWPS) and propose a
mechanism to control a threshold value, i.e., an  activation energy.
In  Sec.~\ref{sec:3}  numerical  results for  SNR, the probability 
density for residence time\cite{12},   stochastic dynamics of threshold values
 and the firing rate
for the TSS are presented. We show that large  SNR is achieved in the 
small noise region as expected.  
 In  Sec.~\ref{sec:4}  the adaptation process, which is governed by 
   a threshold equation with two parameters $\alpha$ and $\beta$,  
 is studied  both analytically and numerically. We discuss  
how these parameters affect  quality of
  information transfer, with main emphasis put on  a large noise region.  
%By choosing proper values for the two
%  parameters,
%we show that information transfer    
 Final section contains some comments  on  energy transfer rate from
 input 
signals to a
 reservoir and   on  
a double well potential
 system(DWPS).  

\section{Model}\label{sec:2}

In this section we first introduce the two-state system(TSS)\cite{11}
  and relate
it to the double-well potential system(DWPS) for convenience for 
later discussions on physical  aspects of the model such as 
energy transfer to a reservoir.
%produced by input signals.%, which  gives a physical
%background of the TSS and .  
The system variable $x(t)$ 
at time $t$ is 
assumed to take only two values, $x_+=1$ and $x_-=-1$ and 
 transition  between the two states is described by the master equation 
\begin{equation}
dp_+(t)/dt=-w_-(t)p_+(t)+w_+(t)(1-p_+(t)), \label{e:1}
\end{equation}
where $p_+(t)$ denotes the probability that $x(t)=x_+$ with $p_+(t)+p_-(t)=1$. 
 $w_-(t)$ is the transition probability at time $t$ for 
the particle to jump to the 
left($x_-$) site and  $w_+(t)$ is similarly defined.
  
The rates $w_+(t),w_-(t)$ are expressed in an Arrhenius form as
\begin{eqnarray}
w_+(t)&=&  \exp[(-a+A_0 \cos(\omega_0 t))/T], \nonumber \\
w_-(t)&=&  \exp[(-a-A_0 \cos(\omega_0 t))/T],\label{e:2}
\end{eqnarray}
where  $T$ measures strength of noise  and 
$a\pm A_0 \cos(\omega_0 t)$ denotes (time dependent) 
activation energy for jumping. 
 
A physical system which is closely related  to the TSS is 
 a double-well potential system(DWPS) described by the 
Langevin equation  
\begin{eqnarray}
dx/dt = -dV(x)/dx+  A_0\cos(\omega_0 t)+f(t), \label{e:3}
\end{eqnarray} 
where the random force $f(t)$ satisfies the fluctuation-dissipation relation
\begin{eqnarray}
\langle f(t)f(t')\rangle= 2T\delta(t-t'),\label{e:4}
\end{eqnarray} 
and $V(x)$ represents the double well potential
\begin{eqnarray}
V(x) &=& a (x-1)^2(x+1)^2.\label{e:5}
\end{eqnarray} 
When  both $A_0$ and $T$ are   smaller than  
 $a$ in Eq.~\eqref{e:5},  Brownian particle described by  
Eq.~\eqref{e:3} may be considered to   
stay either at  $x_+=1$ or $x_-=-1$  for  time of the order of  Kramers time
 $\tau_{Kr}\simeq  \exp(a/T)$\cite{4} %with $K_r=2\sqrt 2 a/\pi$ and 
 and   occasionally  jumps between $x_+$ and $x_-$. 

When the relaxation time $\tau_{r}\simeq (8a)^{-1}$ of intrawell motion 
is short in the
sense  $\tau_r \omega_0 \ll 1$ one can introduce the adiabatic assumption
to reduce the DWPS approximately to a two-state system(TSS) described by 
 Eq.~\eqref{e:1}.% with the hopping rate Eq.~\eqref{e:2}.

Both TSS and DWPS are extensively studied in connection with SR and 
  are known to show SR\cite{1}, that is, SNR shows maximum at nonzero
 $T$ when other parameters charactering the system, such as activation
 energy $a$ and  $\omega_0, A_0$ are kept fixed. It may be noted that 
for the TSS\cite{11,12}  
 analytic (or integral form) results for SNR and the distribution 
function $p_{f.p}(\tau)$ 
of the first passage time for jumping to another state are available.
One merit of the TSS is that even if we take effects of self-tuning(ST) 
into account, 
we can calculate SNR by solving a coupled set of differential equations 
 Eq.~\eqref{e:1} and Eq.~\eqref{e:6}, to be given below,   
without recourse to Monte Carlo simulations, which inevitably introduce
fluctuations to power spectra and consequently to SNR.  
  
Here we introduce a mechanism for  self-tuning(ST) of the
 activation energy $a$
in Eq.~{2}, following the prescription presented in Ref.~\cite{9}.    
If there occurs no jumping or activation events, $a(t)$ simply
decreases, while  
 if a jumping  event occurs $a(t)$ increases, thus controlling the
 jumping
 or  firing rate 
 by  avoiding too large or too small firing rates. 
To express this adaptation process mathematically, we employ the following 
dynamics for $a(t)$,      
\begin{equation}
da(t)/dt=-\alpha a(t) +\beta[w_+(t)p_-(t)+w_-(t)p_+(t)].\label{e:6}
\end{equation}
Indeed, if we tentatively put $\beta=0$, $a(t)$ goes to zero since $\alpha$ is 
chosen to
be positive.  If we put $\beta$ positive, we notice that $a(t)$ increases 
in proportion to the barrier crossing rate. By this mechanism we expect
that the TSS adjusts $a(t)$, reflecting the circumstances it is put in.  

For the DWPS we propose a similar adaptation dynamics for $a(t)$ of the
form
\begin{equation}
da(t)\equiv a(t+dt)-a(t)=-\alpha a(t)dt+\beta \int_t^{t+dt} dt~\Sigma_i\delta(t-t_i),
\label{e:7}
\end{equation}  
where $t_i(i=1,2,...)$ denotes the time  when $x(t)=1$.

\section{Numerical results for TSS}\label{sec:3}

We first explain how one can calculate SNR for the TSS with
self-tuning(ST), by slightly modifying the approach in  Ref.\cite{11}.  

\subsection{SNR with self-tuning: methodology}\label{sec:level2}

Let us denote the solution to Eq.~\eqref{e:1} and Eq.~\eqref{e:6} as 
\begin{equation}
p_+(t)=p_+(t|x_0,a_0,t_0), \hspace{0.6cm} a(t)=a(t|x_0,a_0,t_0),\label{e:8}
\end{equation}
which satisfy the initial conditions
$p_+(t=t_0|x_0,a_0,t_0)=\delta(1,x_0)$ and $a(t=t_0|x_0,a_0,t_0)=a_0$
with $\delta(1,x)$ denoting the Kronecker $\delta$, i.e., $\delta(1,x)=1$ if 
$x=1$ and $\delta(1,x)=0$ if $x\neq 1$.   
The transition
probability $p(x,a,t|x_0,a_0,t_0)$ for $(x(t),a(t))$ to be  
at   $(x,a)$ starting 
from $(x_0,a_0)$ is expressed as
\begin{eqnarray}
p(x,a,t&|&x_0,a_0,t_0)=\delta(a-a(t|x_0,a_0,t_0))\times \nonumber \\
&[& p_+(t|x_0,a_0,t_0)\delta(x-1)\nonumber \\
&+&(1-p_+(t|x_0,a_0,t_0))\delta(x+1)].\label{e:9}
\end{eqnarray}

Following MacNamara and Wiesenfeld\cite{11} let us first introduce 
the time correlation function 
$\phi(t,\tau|x_0,a_0,t_0)$ by
\begin{eqnarray}
\phi(t,\tau&|&x_0,a_0,t_0) = \langle x(t)x(t+\tau)|x_0,a_0,t_0)\rangle \nonumber \\
&\equiv& \int da'\int da
\int dx \int dy xy p(x,a',t+\tau \nonumber \\
&|& y,a,t)p(y,a,t|x_0,a_0,t_0). \label{e:10}
\end{eqnarray}
After performing integration of Eq.~\eqref{e:10} over $y$ and $a$ 
we have
\begin{eqnarray}
\phi(t,\tau&|&x_0,a_0,t_0)=\int da' \int dx x 
[p_+(t|x_0,a_0,t_0)\nonumber \\
&p&(x,a',t+\tau|1,a(t|x_0,a_0,t_0),t)-p_-(t|x_0,a_0,t_0)\nonumber \\
&p&(x,a',t+\tau|-1,a(t|x_0,a_0,t_0),t)].\label{e:11}
\end{eqnarray}
Now we take  the limit 
$t_0\to -\infty$ to remove  $x_0,a_0$ dependence of $p_+, p_-$ and of $a$
on the right hand side of  Eq.~\eqref{e:11}, leading to 
\begin{eqnarray}
\phi(t&,&\tau)=\int da' \int dx x 
[p_+(t)p(x,a',t+\tau|1,a(t),t) \nonumber \\
&-&p_-(t)p(x,a',t+\tau|-1,a(t),t],\label{e:12}
\end{eqnarray}
where we  replace $\lim_{t_0\to -\infty}p_+(t|x_0,a_0,t_0)$ by $p_+(t)$ and 
 $\lim_{t_0\to -\infty }a(t|x_0,a_0,t_0)$ by $a(t)$. 
 $\int da'$ can be performed  trivially  to  have
\begin{eqnarray}
\phi(t,\tau)&=& p_+(t)[2p_+(t+\tau|1,a(t),t)-1]\nonumber \\
   &-&p_-(t)[2p_+(t+\tau|-1,a(t),t)-1]. \label{e:13}
\end{eqnarray}
Finally to make the function $\phi(t,\tau)$ independent of the time 
variable $t$ and also to conform to experimental situations,
we take  time average $(1/\tau_p)\int_0^{\tau_p} dt$ with 
$\tau_p=2\pi/\omega_0$   to obtain
\begin{eqnarray}
\phi(\tau)&=&(1/\tau_p)\int_0^{\tau_p} dt 
\{p_+(t)[2p_+(t+\tau|1,a(t),t)-1]\nonumber \\
   &-&p_-(t)[2p_+(t+\tau|-1,a(t),t)-1\}. \label{e:14}
\end{eqnarray}

Numerical implementation of Eq.~\eqref{e:14} is not difficult and the
result is  conveniently   expressed as 
\begin{eqnarray}
\phi(\tau)\approx\phi_{relax}(\tau)+B\cos(\omega_0\tau), \label{e:15}
\end{eqnarray}
where $\phi_{relax}(\tau)$ is the relaxation part, which
 goes to zero asymptotically as $\tau \to \infty$,
and $B\cos(\omega_0\tau)$ represents the periodic part of the external field.    Fourier transformation
of Eq.~\eqref{e:15} has the form 
\begin{eqnarray}
\tilde \phi(\omega)=\tilde\phi_{relax}(\omega)+
B[\delta(\omega-\omega_0)+\delta(\omega_0+\omega_0)], \label{e:16}
\end{eqnarray}
and  SNR is define here as
\begin{eqnarray}
R_{SN}=B/\tilde\phi_{relax}(\omega_0), \label{e:17}
\end{eqnarray} 

\subsection{Numerical results for SNR and other quantities}\label{sec:level2}

\begin{figure}
\includegraphics[scale=1.5]{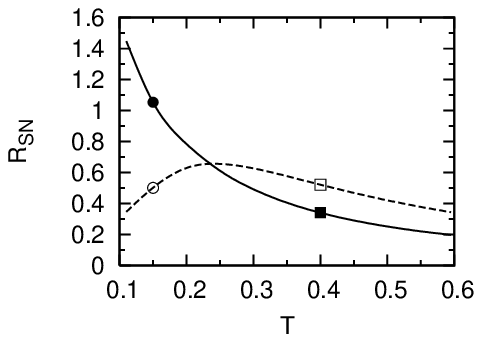}
\caption{
SNR ($R_{SN}$) as a function of noise intensity $T$ for system with(solid curve) 
and without(dashed curve) ST. For ST  we use 
$\alpha=0.03$ and $\beta=0.1$ in Eq.~\eqref{e:6}.  
 The barrier height is set $a=0.5$ for the system without ST.  
Here and hereafter $\omega_0$ and $A_0$ are always set to be 
  $\omega_0=0.5$ and $A_0=0.3$.
}
%\end{figure}
%\begin{figure}
{\bf{a}}

\includegraphics[scale=1.5]{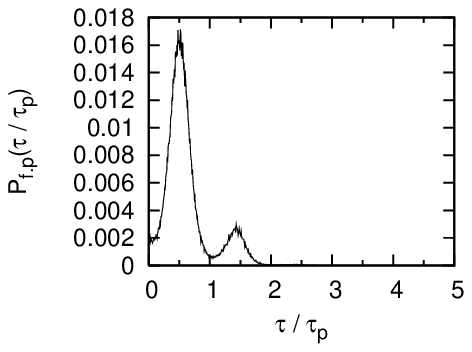}

{\bf{b}}

\includegraphics[scale=1.5]{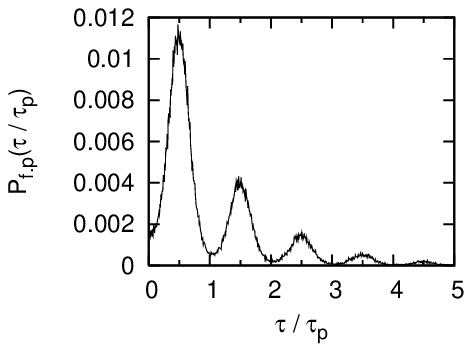}

\caption{
First passage time distribution functions for the system marked by 
black circle(a) and by the white circle(b) in Fig. 1. 
 }
\end{figure}

It is  noted that   
we take 
$\omega_0=0.5$ and $A_0=0.3$  in the following.   
In Fig.1 is plotted SNR for  systems  with self-tuning( 
$\alpha=0.03$ and $\beta=0.1, C\equiv \alpha/\beta=0.3$)
and  without self-tuning($a=0.5$).
 We observe that SNR is improved by
self-tuning in
the low temperature region.
This is confirmed from the first-passage time distribution function 
$p_{f.p}(\tau)$ shown in Fig. 2 for the two systems marked by 
black circle  with  ST and by white circle( without ST) in 
Fig.1 ($T=0.15$).  
 These $p_{f.p}(\tau)$ are obtained by  Monte Carlo simulations,
in which we actually followed particle motion with the hopping rate 
given by   Eq.~\eqref{e:2} and   obtain 
a histogram of the first passage time $\tau$. 
For a system with ST(Fig. 2a) we notice that most of the
particles hop, taking the first chance of low 
activation energy. This is in contrast with the system without
ST(Fig. 2b), for which we observe many bumps of probability with the
spacing $\tau_{p}=2\pi/\omega_0$.\cite{12,13,14}   
\begin{figure}
{\bf{a}}
 \includegraphics[scale=1.5]{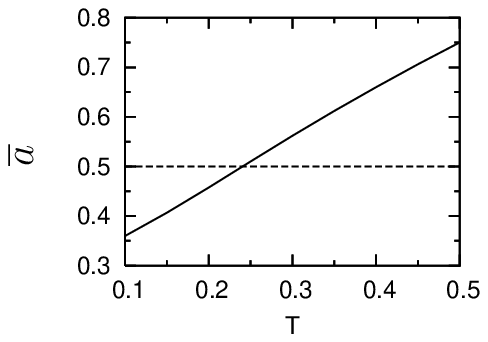}
 {\bf{b}}
 \includegraphics[scale=1.5]{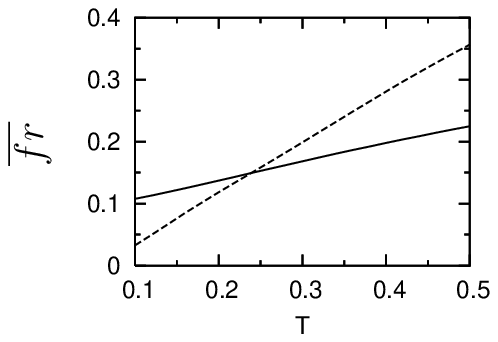}
  \caption{
  Time-averaged activation energy $\overline{a}$  
  (a) and 
  time-averaged firing rate  $\overline{fr}$ as  functions of $T$(b). 
 Parameter values used for the solid and dashed curves correspond 
 to the ones in Fig.1.}
 \end{figure}
\begin{figure}
{\bf{a}}
\includegraphics[scale=1.5]{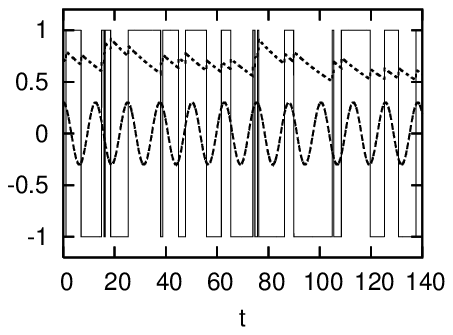}
{\bf{b}}
\includegraphics[scale=1.5]{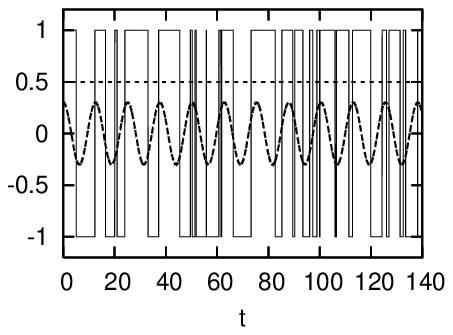}
\caption{ 
Dynamical behavior $x(t)=\pm 1$ (full curves) and $a(t)$(dotted curves) 
from Monte Carlo simulations together with the sinusoidal signals
$A_0\cos(\omega_0 t)$(dashed curves)
 for the system marked by the black squares(a) and
the white squares(b) with $\alpha=0.03$ and $\beta=0.1$. }
\end{figure}
We discuss now the overall $T$-dependence of SNR shown in Fig.1 
based on  
  time-averaged activation energy(Fig. 3a),  
$\overline{a}=\tau_p^{-1}\int_0^{\tau_p} a(t)$
with $\tau_p\equiv 2\pi/\omega_0$ and on a  
time-averaged firing rate $\overline{fr}$ similarly defined 
as $\overline{a}$(Fig. 3b).
In the low temperature region ($T<0.25$), the  firing rate 
$\overline{fr}$ is increased since self-tuning(ST)  lowers 
the activation barrier 
$\overline{a}(<a=0.5)$.
However in the high temperature region($T>0.25$) 
 where noise intensity is high, SNR is deteriorated by ST
 due to  considerable increase of  $\overline{a}$, which results in rapid
decrease of  $\overline{fr}$ compared with the fixed threshold case(Fig.3).
Firing events are in general useful  for information transfer   
and our results suggests that rapid growth of  $\overline{a}$ 
as $T$ increases is not welcome from the point of information processing
by a threshold device. Behavior of $\overline{a}$
and  $\overline{fr}$ depend on the parameters $\alpha$ and $\beta$ in 
 Eq.~\eqref{e:6} and this will be considered in the next section.
 
Before proceeding to this problem, we show a typical 
 example of Monte Carlo trajectories   
 $(x(t),a(t))$ together with the input signal $A_0\cos(\omega_0 t)$
 in Fig.~4a  for the system 
marked by black squares(a) and white squares(b), belonging to 
 a high $T$ region($T=0.4$). 
When $T$  and consequently noise  are large, we have some chances
 of successive hopping events as shown in Fig. 4. In this case 
the activation energy $a(t)$ increases rapidly as shown in Fig. 4a(
 typically around $t\simeq 80$) 
due to ST, which inhibits a firing event on average 
for some time.   That is, in our Monte Carlo simulations 
we increase $a(t)$ by $\beta$ 
 whenever there occurs a hopping event(see Eq. (6)). 
 From this we intuitively see 
that large $\beta$ values makes $a(t)$ large, resulting in small $fr(t)$.
With these preparations we now consider $\alpha$ and $\beta$ dependence 
of SNR.      
 
\section{threshold dynamics and SNR}\label{sec:4}

Now let us consider Eq.~\eqref{e:6}, which describes time evolution of
the barrier height $a(t)$, and   express  it  as 
\begin{eqnarray}
da/dt=-\alpha a(t)+\beta fr(t), \label{e:18}
\end{eqnarray}            
where $fr(t)$ denotes a firing rate at time $t$. 
 Since we are mainly 
interested in a  subthreshold situation (i.e. $a(t)>A_0$) 
and a large $T$ region where  
ST  did not work well
 compared with the weak noise region(see Fig.1), we  neglect for 
  qualitative discussion     
$A_0/T$ in  Eq.~\eqref{e:2} and  obtain, with a use of simple form for
 Kramers rate\cite{4},    
\begin{eqnarray}
(\alpha/\beta)\overline{a}\equiv C\overline{a}=
\exp(-\overline{a}/T), \label{e:19}
\end{eqnarray}   
\begin{figure}[hbp!]
{\bf{a}}
\includegraphics[scale=1.5]{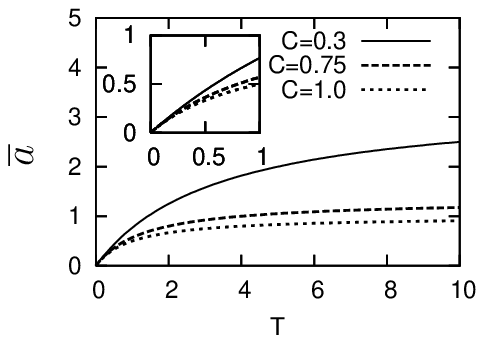}
{\bf{b}}
\includegraphics[scale=1.5]{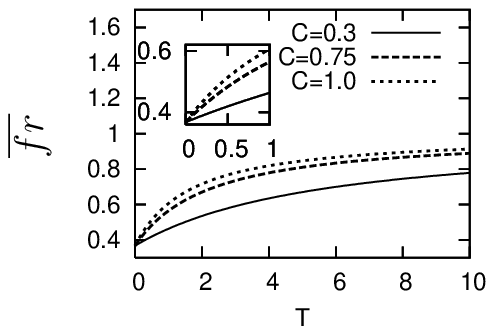}
\caption{
 Time averaged activation energy $\overline{a}$(a) and 
 time-averaged firing rate  $\overline{fr}$(b) as  functions of $T$ 
 from Eq. (19)}
\end{figure}
\begin{figure}
\includegraphics[scale=1.5]{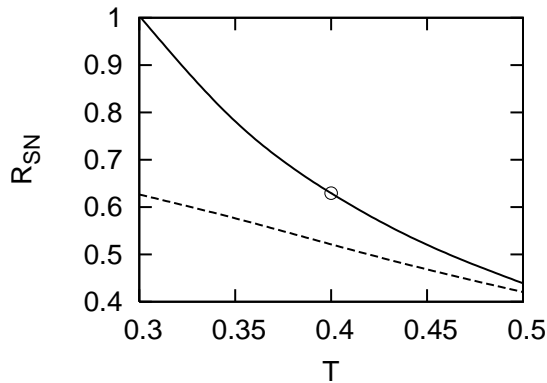}
\caption{
SNR ($R_{SN}$) with  ST($\alpha=0.03, \beta=0.04, C=0.75$)
(full curve) and without ST(dashed curve).}
\end{figure}
 after  time averaging  both
sides of   Eq.~\eqref{e:18} for one period $\tau_p=2\pi/\omega_0$
 of the external field.  
From this we see that at large 
$T$, $\overline{a}\to 1/C$ and 
$\overline {fr}\to (1-1/(TC))$. 
 In Fig. 5a 
we show $\overline{a}$  as a function of $T$, 
 which is obtained by solving
 Eq.~\eqref{e:19}  for three values of 
$C\equiv \alpha/\beta$($C=0.3,0.75, 1.0$ from above).  
We  notice that the barrier height
$\overline{a}$ remains small even for large $T$  when $C\equiv
\alpha/\beta$
 becomes large. 
The firing rate  $\overline{fr}$, calculated from  Eq.~\eqref{e:2}
,  Eq.~\eqref{e:6}, and 
  Eq.~\eqref{e:19}, is shown  in Fig. 5b($C=0.3,0.75, 1.0$ from below). 
Reflecting the fact that  $\overline{a}$ does not increase rapidly with
$T$ when $C$ is large, 
 the firing rate seems to remain large in a large $T$ region when   
 $C$ becomes slightly larger than 0.3.

Guided by this observation we choose $C=0.75$( 
$\alpha=0.03$ and $\beta=0.04$) and plot SNR in Fig. 6  as a function of
$T$. % SNR larger than 1. (no problem?)
Compared with the solid curve in Fig. 1 we notice that SNR is improved 
considerably and our  ST seems
to work well even in the high $T$ region by choosing proper values for
%$\alpha$ and $\beta$ and consequently for
 $C=\alpha/\beta$.

\begin{figure}
\includegraphics[scale=1.5]{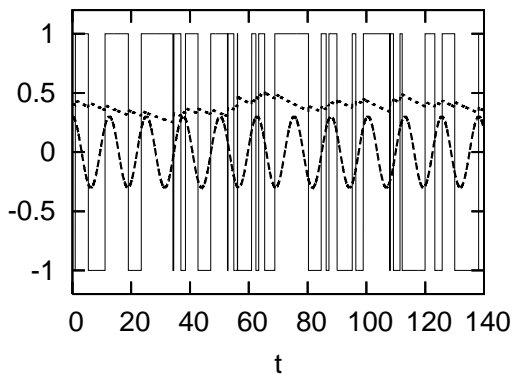}
\caption{
Dynamical behavior $x(t)=\pm 1$ (full curves) and $a(t)$(dotted curves) 
from Monte Carlo simulations together with the sinusoidal signals
$A_0\cos(\omega_0 t)$(dashed curves) for the system marked by the white circle in Fig. 6.
}
\end{figure}

Details of dynamics $(x(t),a(t))$ are shown in Fig. 7 for the system
marked with a white circle in Fig. 6. This should be compared with the
dynamics in Fig. 4a which is characterized by different 
parameter values($\alpha=0.03,\beta=0.1,C=0.3$).  By choosing a smaller
value for $\beta(=0.04)$(keeping $\alpha$ fixed to 0.03) 
 we could prevent the activation energy becoming too large
and this contributes to making SNR large. 
%\section{energy transfer  and conclusion}\label{sec:5}
% \begin{figure}
% 
% {\bf{a}}
% 
% \includegraphics[scale=1.5]{fig5a.eps}
% 
% {\bf{b}}
% 
% \includegraphics[scale=1.5]{fig5b.eps}
% 
% \caption{
%  Time averaged activation energy $\overline{a}$(a) and 
%  time-averaged firing rate  $\overline{fr}$(b) as  functions of $T$ 
%  from Eq. (19)}
% \end{figure}
% 
% \begin{figure}
% \includegraphics[scale=1.5]{fig6.eps}
% \caption{
% SNR ($R_{SN}$) with  ST($\alpha=0.03, \beta=0.04, C=0.75$)
% (full curve) and without ST(dashed curve).}
% \end{figure}
% \begin{figure}
% \includegraphics[scale=1.5]{fig7.eps}
% \caption{
% Dynamical behavior $x(t)=\pm 1$ (full curves) and $a(t)$(dotted curves) 
% from Monte Carlo simulations together with the sinusoidal signals
% $A_0\cos(\omega_0 t)$(dashed curves) for the system marked by the white circle in Fig. 6.
% }
% \end{figure}

\section{energy transfer, DWPS  and conclusion}\label{sec:5}

In this section we consider briefly energy transfer from  input signals
to the reservoir(i.e. dissipation) and the DWPS, Eqs. (3-5) before
concluding this paper. 

The hopping rate, Eq. (6), can be rewritten as 
\begin{eqnarray}
w_+(t)&=&  \exp[-(V_s-V_1(t))/T], \nonumber \\
w_-(t)&=&  \exp[-(V_s-V_{-1}(t))/T],
\end{eqnarray}   
with $V_s(=a)$  and $V_{\pm 1}(t)$ the energy at the saddle point($x=0$)
 and at the position $x=\pm 1$, respectively.
If $x(t)$ changes at $t=t_1$ from -1 to 1, the energy $\Delta E(t_1)$ 
transfered from the signal to the reservoir  
  is given by $\Delta E=-(V_1(t_1)-V_{-1}(t_1))=2V_{-1}(t_1)$. 
Dividing all the energy  $\sum_{i} \Delta E(t_i)$ by
 the experimental duration
 $\tau_{exp}$ and $\overline{a}$, we have 
\begin{eqnarray}
E_{s \to r}=\sum_{i} \Delta E(t_i)/(\tau_{exp}\overline{a}),\label{e:21}
\end{eqnarray}    
which was obtained by Monte Carlo experiments. 
\begin{figure}[hbp!]
\includegraphics[scale=1.5]{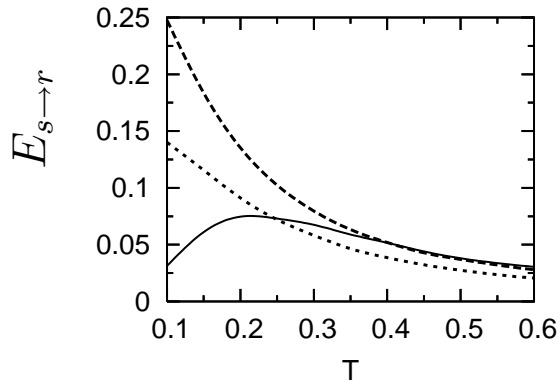}
\caption{
 Energy transfer rate $E_{s \to r}$ as a function of $T$
 for s system without ST(full curve) and with ST with  
$\alpha=0.03,\beta=0.1$ (dotted curve) and $\alpha=0.03,\beta=0.04$
(dashed curve).} 
\end{figure}
\begin{figure}
\includegraphics[scale=1.5]{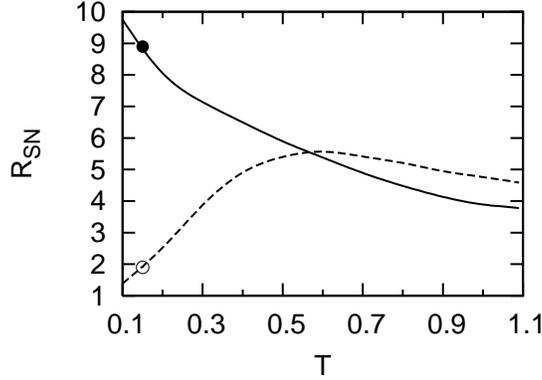}
\caption{
SNR ($R_{SN}$) for DWPS with ST(full curve) and 
without ST(dashed curve). For ST we use $\alpha=0.05,~\beta=0.05$ in
 Eq.~\eqref{e:7} and the barrier height is set $a=1$ for the system without
 ST where $A_0=0.8$ and $\omega_0=0.5$.}
% First passage time distribution function $p_{f.p}(\tau)$ 
%for DWPS with ST() 
\end{figure}

\begin{figure}
{\bf{a}}
\includegraphics[scale=1.5]{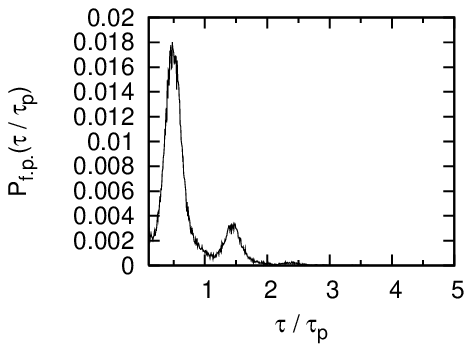}
{\bf{b}}
\includegraphics[scale=1.5]{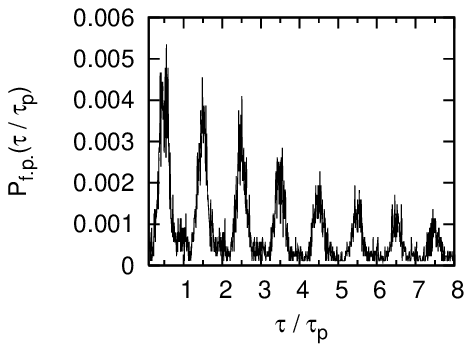}
\caption{  First passage time distribution function $p_{f.p}(\tau/\tau_p)$ 
for DWPS with ST(a) and without ST(b) at $T=0.15$. Parameter values
 characterizing the system is the same with those for Fig. 9. }
\end{figure}

In Fig. 8 is plotted $E_{s \to r}$ as a function of $T$. The dotted
curve($\alpha=0.03,\beta=0.04$), the dashed
curve($\alpha=0.03,\beta=0.1$), and the full curve correspond to the
systems represented by the full curve in Fig. 6, the full curve in Fig. 1 and the dashed curve
in Fig. 1, respectively. We see that SNR and $E_{s \to r}$ show
surprisingly similar behaviors. This is rather natural since both
quantities depend on the firing rate and the firing timing in similar
ways. Especially the firing timing is important for both SNR and 
$E_{s \to r}$. 
When  a hopping event from $x=-1$ to $x=1$ occurs at time $t_1$,
 maximum energy transfer is achieved when $V_{-1}$ becomes maximum at 
time $t_1$. This synchrony is evidently reflected  to SNR. As noted in
Sec.~I the synchrony is also important for SNR.

Final comment is on the double-well potential system(DWPS), Eqs. (3-5).
Since TSS and DWPS describe similar hopping events under time
 periodic signals, we expect that both systems share common properties 
, especially in relation to  ST and SR. Fig. 9 shows SNR of DWPS
with(full curve) and without(dashed curve) ST, where SNR
is defined as the ratio 
$P(\omega_0)/[P(\omega_0-d\omega)/2+P(\omega+d\omega)/2]$
with $P(\omega)$ denoting the power spectral density at  frequency
$\omega$ and $d\omega$ is the frequency mesh size in numerical calculations of $P(\omega)$. This should be compared with 
Fig.1 for TSS. Corresponding to Fig. 2, we compare  $p_{f.p}(\tau)$
  for the two systems marked by a white and black circle in Fig. 9. in Fig.10. 
From these results it is seen that TSS and DWPS behave similarly 
with respect to response to and information transfer of the periodic signals.

In this paper we applied a ST mechanism, Eq.~\eqref{e:6}, to TSS, Eq.~\eqref{e:1} and 
confirmed that better SNR is simply obtained by ST mechanism for  small
fluctuation(i.e. low $T$) region.  Tuning of the parameters
$\alpha$ and $\beta$ was guided by a simple equation~\eqref{e:19}, leading to
better SNR  even for a high $T$ region. Energy transfer or dissipation
rate was also studied and this quantity ~\eqref{e:21} turned out to be able to
 play a similar role as a measure for information processing ability
of a threshold device.   
 
%\newpage


\begin{thebibliography}{0}

\bibitem{1}  V. S. Anishchenko, V. V. Astakhov, A. B. Neiman, 
T. E. Vadivasova, and L. Schimansky-Geier,
{\it Nonlinear Dynamics of Chaotic and Stochastic Systems}, 
(Springer, Berlin, 2002).  
  L. Gammaitoni, P. H\"{a}nggi, P. Jung, and F. Marchesoni,
 Rev. Mod. Phys. {\bf 70},  223 (1998).
%physical
\bibitem{2} P. Reimann, Phys. Rep. {\bf 361}, 57 (2002).
F. J\"{u}licher, A. Ajdari, and J. Prost, Rev. Mod. Phys. {\bf 69}, 
 1269 (1997).
%biological
\bibitem{3} S.Kirkpatrick, C. D. Gelatt, and M. P. Vecchi, Science
{\bf 220}, 671 (1983).
%technological
\bibitem{4} H. A. Kramers, Physica {\bf 7}, 284 (1940). 
S. Chandrasekhar, Rev. Mod. Phys. {\bf 15}, 1 (1943), in 
{\it Selected Papers on Noise and Stochastic Processes}, edited by
	N. Wax, (Dover Publications, New York 1954).
%Kramers time
\bibitem{5} A. K. Hartmann, and H. Rieger, {\it Optimization Algorithms
	in Physics}, (Wiley-VCH, Berlin 2002).
\bibitem{6} F. Marchesoni, F. Apostolico, and S. Santucci, Phys. Rev. E
		{\bf 59}, 3958 (1999). F. Apostolico, L. Gammaitoni,
		F. Marchesoni, and S. Santucci, Phys. Rev. E {\bf 55}, 36 (1997).
% threshold system
\bibitem{7} R. Benzi, S. Sutera, and A. Vulpiani, J. Phys. A {\bf 14},
 L453 (1981). 
\bibitem{8} A. B. Neiman, D. F. Russel, X. Pei, W. Wojtenek, J. Twitty,
	E. Simonotto, B. A. Wetting, E. Wagner, L. A. Wilkens, F. Moss,
 Int. J. Bifurcation Chaos, {\bf 10},
2499 (2000).
\bibitem{9} T. Munakata, T. Hada, and M. Ueda, Physica A. {\bf 375},
  492 (2007). 
\bibitem{10} W. Denk, W. W. Webb, A. J. Hudspeth, Proc. Natl. Acad. Sci.
USA {\bf 86}, 5371 (1989). 
S. Camalet, T. Duke, F. J\"{u}licher, J. Prost, Proc. Natl. Acad. Sci.
USA {\bf 97}, 3183 (2000).  
\bibitem{11} B. McNamara, and K. Wiesenfeld, Phys. Rev. A. {\bf  39},
 4854 (1989).
\bibitem{12}T. Zhou, F. Moss, and P. Jung, Phys. Rev. A {\bf 42}, 3161 (1990).
\bibitem{13} L. Gammaitoni, F. Marchesoni, E. Menichella-Saetta, 
and S. Santucci, Phys. Rev. Lett. {\bf 62}, 349 (1989).
\bibitem{14} L. Gammaitoni, F. Marchesoni, and S. Santucci,
		Phys. Rev. Lett. {\bf 74}, 1052 (1995).
\end{thebibliography}
\end{document}